\documentclass[pra,preprint,superscriptaddress,amsmath,amssymb]{revtex4}
\usepackage{graphicx}
\usepackage{bm}
\usepackage{amsmath,amssymb,theorem}
\theorembodyfont{\normalfont}
\newtheorem{theorem}{Theorem}
\newtheorem{lemma}[theorem]{Lemma}
\newtheorem{definition}[theorem]{Definition}

\begin{document}
\title{Impossibility of blind quantum sampling 
for classical client}
YITP-18-119
\author{Tomoyuki Morimae}
\email{tomoyuki.morimae@yukawa.kyoto-u.ac.jp}
\affiliation{Yukawa Institute for Theoretical Physics,
Kyoto University, Kitashirakawa Oiwakecho, Sakyo-ku, Kyoto 606-8502, Japan}
\affiliation{JST, PRESTO, 4-1-8 Honcho, Kawaguchi, Saitama 332-0012, Japan}
\author{Harumichi Nishimura}
\email{hnishimura@is.nagoya-u.ac.jp}
\affiliation{Graduate School of Informatics, Nagoya University,
Furocho, Chikusaku, Nagoya, Aichi, 464-8601, Japan}
\author{Yuki Takeuchi}
\email{takeuchi.yuki@lab.ntt.co.jp}
\affiliation{NTT Communication Science Laboratories,
NTT Corporation, 3-1 Morinosato Wakamiya, Atsugi, Kanagawa
243-0198, Japan}
\author{Seiichiro Tani}
\email{tani.seiichiro@lab.ntt.co.jp}
\affiliation{NTT Communication Science Laboratories,
NTT Corporation, 3-1 Morinosato Wakamiya, Atsugi, Kanagawa
243-0198, Japan}

\begin{abstract}
Blind quantum computing enables a client,
who can only generate or measure single-qubit states,
to delegate quantum computing to a remote quantum server
in such a way that the input, output, and program are hidden
from the server. It is an open problem whether a completely classical
client can delegate quantum computing blindly.
In this paper, we show that if a completely classical client can
blindly delegate sampling of subuniversal models,
such as the DQC1 model and the IQP model,  
then the polynomial-time hierarchy collapses to the third level.
Our delegation protocol is the one where the client first
sends a polynomial-length bit string to the server
and then the server returns a single bit to the client.
Generalizing the no-go result to more general setups is an open problem. 
\end{abstract}

\date{\today}
\maketitle  

\section{Introduction}
Blind quantum computing~\cite{BFK,Barz,BarzNP,Chiara,Vedrancomposability,FK,Vedran,MABQC,HayashiMorimae,AKLTblind,topoblind,CVblind,Lorenzo,Atul,Sueki,Takeuchi,distillation} enables a client (Alice),
who can access to only a limited quantum technology,
to delegate her quantum computing to a remote quantum server (Bob)
in such a way that Alice's input, output, and program are hidden
from Bob.
The first blind quantum computing protocol proposed by
Broadbent, Fitzsimons, and Kashefi~\cite{BFK} requires
Alice to do only preparations of randomly-rotated
single-qubit states.
It is open whether the requirement is removed: it is open
whether a completely classical Alice can blindly delegate quantum computing
to Bob.
Several other blind quantum computing protocols have been proposed to ease
Alice's burden.
For example, generating weak coherent pulses
was shown to be enough instead of the single-qubit state
generation~\cite{Vedran}.
Furthermore, it was shown that blind quantum computing is possible
for Alice who can do only single-qubit measurements~\cite{MABQC}. 
Measuring quantum states is sometimes easier than
generating quantum states.
However, all previous results require some minimum quantum
technologies for Alice, and 
the possibility of blind quantum computing for
completely classical Alice remains open.
(Note that we are interested in the 
information-theoretic security. If we consider the computational one,
a recent breakthrough showed that
secure delegated quantum computing for completely classical Alice
is possible~\cite{Mahadev}.)

Morimae and Koshiba showed that
if one-round perfectly-secure
delegated universal quantum computing is possible for a completely classical
client, then
${\rm BQP}\subseteq{\rm NP}$~\cite{MorimaeKoshiba}.
Since BQP is not believed to be in NP, the result suggests
the impossibility of such a 
delegation. In their protocol, only a single round of message exchange
is done between Alice and Bob, and what is sent from
Bob to Alice is only a single bit.
Aaronson, Cojocaru, Gheorghiu, and Kashefi considered a more
general setup where
poly-length messages are exchanged in poly-rounds
between Alice and Bob~\cite{ACGK,error}. 

In this paper, we consider a classical blind delegation
of sampling of subuniversal models. We show that the delegation is
impossible unless the polynomial-time
hierarchy collapses to the third level.
The result holds for any subuniversal
model that does not change
the complexity class NQP (or that is universal under postselections,
see Sec.~\ref{sec:generalizations}).
Examples are the DQC1 model~\cite{KL}, the IQP model~\cite{BJS},
the depth-four model~\cite{TD}, the Boson Sampling model~\cite{AA}, 
the random circuit model~\cite{random}, 
and the HC1Q model~\cite{HC1Q}, etc.
In our protocol, only a single-round of message exchange is done
and what Bob sends to Alice is only a single bit.
It is an open problem whether our no-go result is generalized
to more general setups (for discussion on this point,
see Sec.~\ref{sec:discussion}).
One might think that in the case when Bob sends only 
a single bit to Alice, a no-go result could be shown
unconditionally. However, some computational assumptions
seem to be necessary.
In fact, the blind delegation of
BPP sampling can be done unconditionally: Alice has only to
do it by herself.

This paper is organized as follows.
In the next section,
Sec.~\ref{sec:preliminaries},
we give some preliminaries.
In Sec.~\ref{sec:setup}, we explain the delegation
protocol we consider.
In Sec.~\ref{sec:result}, we show the no-go result for
the DQC1 model.
In Sec.~\ref{sec:generalizations}, we 
generalize the result to other subuniversal
models.
Finally, in Sec.~\ref{sec:discussion},
we give some discussions.

\section{Preliminaries}
\label{sec:preliminaries}
In this section, we provide some preliminaries
necessary to understand the main result.
\subsection{DQC1}
Let us first explain the DQC1 model.
The DQC1 model is a restricted model of quantum computing
where all but a single input qubit are maximally mixed.
It was introduced by Knill and Laflamme originally
to model NMR quantum computing~\cite{KL}.
It is known that the DQC1 model can efficiently solve several problems
whose classical efficient solutions are not known,
such as the calculation of Jones polynomials~\cite{ShorJordan}.
Furthermore, it was shown that output probability distributions
of the DQC1 model cannot be classically efficiently sampled
unless the polynomial-time hierarchy 
collapses~\cite{DQC1additive,DQC1nonclean,DQC1_1,DQC1_1ICALP,DQC1_3}.

Let $V$ be a quantum circuit on $n$ qubits.
We define the probability distribution 
$p_V^{DQC1}:\{0,1\}\to [0,1]$ by
\begin{eqnarray*}
p_V^{DQC1}(0)&=&\mbox{Tr}\Big[(|0\rangle\langle0|\otimes I^{\otimes n-1})
V\Big(|0\rangle\langle0|\otimes\frac{I^{\otimes n-1}}{2^{n-1}}\Big)
V^\dagger
\Big],\\
p_V^{DQC1}(1)&=&\mbox{Tr}\Big[(|1\rangle\langle1|\otimes I^{\otimes n-1})
V\Big(|0\rangle\langle0|\otimes\frac{I^{\otimes n-1}}{2^{n-1}}\Big)
V^\dagger
\Big],
\end{eqnarray*}
where $I\equiv|0\rangle\langle0|+|1\rangle\langle1|$ is the
two-dimensional identity operator.
In this paper we consider the classical delegation of sampling of
$\{p_V^{DQC1}(z)\}_{z\in\{0,1\}}$.
It is known that if 
$\{p_V^{DQC1}(z)\}_{z\in\{0,1\}}$ is sampled
in classical polynomial time with a multiplicative error
$0\le \epsilon<1$, then the polynomial-time
hierarchy collapses to the second level~\cite{DQC1_1,DQC1_1ICALP}.
Here, we say that a probability distribution $\{p_z\}_z$ is
sampled in classical polynomial time with a multiplicative error
$\epsilon$ if there exists a classical probabilistic
polynomial-time algorithm that outputs $z$ with probability $q_z$
such that
\begin{eqnarray*}
|p_z-q_z|\le\epsilon p_z
\end{eqnarray*}
for all $z$.

\subsection{IQP}
We next explain the
IQP model~\cite{BJS,BMS}. An $n$-qubit IQP circuit is
a quantum circuit in the form of
$H^{\otimes n}UH^{\otimes n}$, where $H$ is an Hadamard gate and
$U$
is a quantum circuit that consists of only
$Z$-diagonal gates, such as $Z$, $CZ$, $CCZ$, and 
$e^{i\theta Z}$.
Here
\begin{eqnarray*}
Z&\equiv&|0\rangle\langle0|-|1\rangle\langle1|,\\
CZ&\equiv&I^{\otimes 2}-2|11\rangle\langle11|,\\
CCZ&\equiv&I^{\otimes 3}-2|111\rangle\langle111|.
\end{eqnarray*}

Let $V$ be an $n$-qubit IQP circuit.
For an $m\le n$,
we define the probability distribution
$p_V^{IQP,m}:\{0,1\}\to[0,1]$ by
\begin{eqnarray*}
p_V^{IQP,m}(1)&=&\big\|(|1^m\rangle\langle1^m|\otimes I^{\otimes n-m})
V|0^n\rangle\big\|^2,\\
p_V^{IQP,m}(0)&=&1-p_V^{IQP,m}(1).
\end{eqnarray*}
In this paper we consider the classical delegation of 
sampling of $\{p_V^{IQP,m}(z)\}_{z\in\{0,1\}}$
with $m=poly(n)$.
It is known that if
$\{p_V^{IQP,m}(z)\}_{z\in\{0,1\}}$ is sampled for certain $m=poly(n)$ in
classical polynomial time with a multiplicative error
$0\le \epsilon<1$,
then the polynomial-time hierarchy collapses to the 
second level~\cite{DQC1_1,DQC1_1ICALP}.
(It is also known that $\{p_V^{IQP,m}(z)\}_z$ is 
exactly sampled in classical polynomial time if 
$m=O(\log(n))$~\cite{BJS}.)

\subsection{NQP}
The complexity class NQP is a quantum version of NP and
defined as follows~\cite{ADH97}:

\begin{definition}
A problem 
$A=(A_{yes},A_{no})$ is in NQP if and only if
there exists a polynomial-time uniformly generated family $\{V_x\}_x$
of quantum circuits such that
\begin{itemize}
\item
If $x\in A_{yes}$ then $p_{V_x}(1)>0$.
\item
If $x\in A_{no}$ then $p_{V_x}(1)=0$.
\end{itemize}
Here,
$
p_{V_x}(1)\equiv\langle 0^n|V_x^\dagger
(|1\rangle\langle1|\otimes I^{\otimes n-1})
V_x|0^n\rangle$,
and $n=poly(|x|)$.
\end{definition}

It is important to point out that quantum computing
in the above definition of NQP
can be restricted to some subuniversal models, such as
the DQC1 model or the IQP 
model~\cite{DQC1_1,DQC1_1ICALP}.
Let us define the following classes.

\begin{definition} 
A problem
$A=(A_{yes},A_{no})$ is in ${\rm NQP}_{\rm DQC1}$ 
if and only if
there exists a polynomial-time uniformly generated family $\{V_x\}_x$
of quantum circuits such that
\begin{itemize}
\item
If $x\in A_{yes}$ then $p_{V_x}^{DQC1}(1)>0$.
\item
If $x\in A_{no}$ then $p_{V_x}^{DQC1}(1)=0$.
\end{itemize}
Here,
\begin{eqnarray*}
p_{V_x}^{DQC1}(1)\equiv{\rm Tr}
\Big[(|1\rangle\langle1|\otimes I^{\otimes n-1})
V_x\Big(|0\rangle\langle0|\otimes\frac{I^{\otimes n-1}}{2^{n-1}}
\Big)V_x^\dagger\Big],
\end{eqnarray*}
and $n=poly(|x|)$.
\end{definition}

\begin{definition}
A problem $A=(A_{yes},A_{no})$ is in ${\rm NQP}_{\rm IQP}$ if
and only if there exists a polynomial-time uniformly generated 
family $\{V_x\}_x$ of
IQP circuits 
such that
\begin{itemize}
\item
If $x\in A_{yes}$ then $p_{V_x}^{IQP,m}(1)>0$.
\item
If $x\in A_{no}$ then $p_{V_x}^{IQP,m}(1)=0$.
\end{itemize}
Here $p_{V_x}^{IQP,m}(1)\equiv\|(|1^m\rangle\langle1^m|\otimes
I^{\otimes n-m})V_x|0^n\rangle\|^2$,
$n=poly(|x|)$, and $m\le n$. 
\end{definition}

Then we can show the following equivalences.
\begin{theorem}\cite{DQC1_1,DQC1_1ICALP}
${\rm NQP}={\rm NQP}_{\rm DQC1}$.
\label{thm:NQP_DQC1}
\end{theorem}

\begin{theorem}\cite{DQC1_1,DQC1_1ICALP}
${\rm NQP}={\rm NQP}_{\rm IQP}$.
\label{thm:NQP_IQP}
\end{theorem}
For the convenience of readers,
their proofs are given in Appendix~\ref{app:NQP_DQC1}
and Appendix~\ref{app:NQP_IQP}, respectively.

\subsection{${\rm \widehat{BP}}$ operator}
Let K be a complexity class.
The class $\widehat{\rm BP}\cdot {\rm K}$ is defined
as follows~\cite{TO92}.

\begin{definition}
A problem $A=(A_{yes},A_{no})$ is in 
$\widehat{\rm BP}\cdot {\rm K}$ if and only if for any polynomially
bounded function $q:{\mathbb N}\to {\mathbb N}$, there exist
a problem $B=(B_{yes},B_{no})$ in ${\rm K}$ and a polynomially
bounded function $r:{\mathbb N}\to {\mathbb N}$ such that for
every $x\in \{0,1\}^*$, it holds that
\begin{itemize}
\item
If $x\in A_{yes}$ then
\begin{eqnarray*}
|\{z\in\{0,1\}^{r(|x|)}~|~\langle x,z\rangle\in B_{yes}\}|\ge
(1-2^{-q(|x|)})2^{r(|x|)}.
\end{eqnarray*}
\item
If $x\in A_{no}$ then
\begin{eqnarray*}
|\{z\in\{0,1\}^{r(|x|)}~|~\langle x,z\rangle\in B_{no}\}|\ge
(1-2^{-q(|x|)})2^{r(|x|)}.
\end{eqnarray*}
\end{itemize}
\end{definition}

\subsection{Advice classes}
We also use advice classes~\cite{KarpLipton}. 
Let K be any complexity class.
\begin{definition}
A problem $A=(A_{yes},A_{no})$ is in ${\rm K}/{\rm poly}$
if and only if there exist a problem $B=(B_{yes},B_{no})$ in K,
an advice function $f:{\mathbb N}\to\{0,1\}^*$,
and a polynomial $p$ such that $|f(n)|\le p(n)$ for all $n$,
and
\begin{itemize}
\item
If $x\in A_{yes}$ then $\langle x,f(|x|)\rangle\in B_{yes}$.
\item
If $x\in A_{no}$ then $\langle x,f(|x|)\rangle\in B_{no}$.
\end{itemize}
\end{definition}

We also use a complexity class with probabilistic
advice.
\begin{definition}
A problem $A=(A_{yes},A_{no})$ is in
${\rm NP}/{\rm rpoly}$ if and only if
there exist a classical 
probabilistic polynomial-time algorithm $M$ and 
a family $\{q_s\}_{s\in{\mathbb N}}$ of
probability distributions
$q_s:\{0,1\}^{poly(s)}\to[0,1]$ such that
\begin{itemize}
\item
If $x\in A_{yes}$ then 
${\rm Pr}(\mbox{$M$ accepts})>0$.
\item
If $x\in A_{no}$ then 
${\rm Pr}(\mbox{$M$ accepts})=0$.
\end{itemize}
Here, $M$ takes $(x,b)$ as the input,
where $b\in\{0,1\}^{poly(|x|)}$
is sampled from the probability distribution $q_{|x|}$.
\end{definition}

\section{Delegation protocol}
\label{sec:setup}
In this section we explain the delegation protocol we consider.
Let $V_x$ be a quantum circuit on $n$ qubits with a parameter
$x\in\{0,1\}^*$.
Alice wants to sample
the probability distribution
$\{p_{V_x}^{DQC1}(z)\}_{z\in \{0,1\}}$,
where
\begin{eqnarray*}
p_{V_x}^{DQC1}(z)\equiv{\rm Tr}\Big[
(|z\rangle\langle z|\otimes I^{\otimes n-1})
V_x\Big(|0\rangle\langle0|\otimes\frac{I^{\otimes n-1}}{2^{n-1}}
\Big)V^\dagger_x\Big].
\end{eqnarray*}
However, she is completely classical (i.e., her computational
ability is classical probabilistic polynomial-time), so she delegates
the sampling to Bob.
Bob's computational ability is unbounded.
She wants to hide the parameter $x$ from Bob up to its size $|x|$.

Our delegation protocol consists of the following elements:
\begin{itemize}
\item
A classical probabilistic polynomial-time
key generation algorithm $K$.
On input $x\in\{0,1\}^*$, 
the algorithm $K$ outputs $k\leftarrow K(x)$ 
with certain probability, 
where $k\in\{0,1\}^{poly(|x|)}$ is a key.
\item
A classical deterministic polynomial-time
encryption algorithm $E$.
On input $(x,k)\in\{0,1\}^*\times\{0,1\}^{poly(|x|)}$,
it outputs
$a=E(x,k)$, where $a\in\{0,1\}^{poly(|x|)}$.  
\item
A classical deterministic polynomial-time
decryption algorithm $D$.
On input $(x,k,b)\in\{0,1\}^*\times\{0,1\}^{poly(|x|)}
\times\{0,1\}$, it outputs
$\tau=D(x,k,b)$, where $\tau\in\{0,1\}$.
\end{itemize}
Our delegation protocol runs as follows:
\begin{itemize}
\item[1.]
On input $x\in\{0,1\}^*$,
Alice runs the key generation algorithm $K$ to
get a key $k$.
\item[2.]
Alice computes $a=E(x,k)$,
and sends $a$ to Bob.
\item[3.]
Bob sends Alice $b\in\{0,1\}$ with probability
$q_a(b)$.
\item[4.]
Alice computes $\tau=D(x,k,b)$. 
\end{itemize}

We require that this delegation protocol satisfies both the correctness and
blindness simultaneously. Here,
the correctness and blindness are defined as follows.
\begin{definition}
We say that the above protocol is $\epsilon$-correct
if for any circuit $V_x$, any parameter $x$,
and any key $k$ for $x$,
i.e., $k\leftarrow K(x)$,
\begin{eqnarray*}
\big|\mbox{Pr}(\tau=0)-p_{V_x}^{DQC1}(0)\big|
&\le&\epsilon p_{V_x}^{DQC1}(0),\\
\big|\mbox{Pr}(\tau=1)-p_{V_x}^{DQC1}(1)\big|
&\le&\epsilon p_{V_x}^{DQC1}(1).
\end{eqnarray*}
Here,
\begin{eqnarray*}
{\rm Pr}(\tau=z)\equiv\sum_{b\in\{0,1\}}q_{E(x,k)}(b)
\delta_{z,D(x,k,b)}
\end{eqnarray*}
for $z\in\{0,1\}$.
\end{definition}
It means that
Alice can sample $\{p_{V_x}^{DQC1}(z)\}_{z\in\{0,1\}}$
with a multiplicative error $\epsilon$.

\begin{definition}
\label{def:blindness}
We say that the above protocol is blind if the following is satisfied:
Let $x_1$ be any parameter, and
$k_1$ be any key for $x_1$, i.e.,
$k_1\leftarrow K(x_1)$.
For any parameter $x_2$ such that $|x_2|=|x_1|$, 
the probability that $K(x_2)$ outputs $k_2$
such that $E(x_2,k_2)=E(x_1,k_1)$
is non-zero.
\end{definition}
In other words, let $P_x(a)$ be the probability
that $K(x)$ outputs
$k$ such that $a=E(x,k)$. 
Then, the blindness means that supports of
$P_{x_1}$ and $P_{x_2}$ are the same for any
$x_1$ and $x_2$.
The intuition behind Definition~\ref{def:blindness} is that
if such $k_2$ is never generated then
Bob can learn that Alice's parameter is not $x_2$ when
he receives $E(x_1,k_1)$ from Alice.

To conclude this section, we provide several remarks.
First, our delegation protocol is similar to the generalized
encryption scheme (GES) of Refs.~\cite{AFK,ACGK}.
However, the GES is a protocol that enables a client to delegate
the calculation of $f(x)$ for a function $f$ and input $x$
with success probability larger than $1/2+1/poly(|x|)$.
The computation of
the value of $p_{V_x}^{DQC1}(1)$
could be delegated by using
the GES,
but it is stronger than what Alice wants to do, i.e., 
the sampling of $\{p_{V_x}^{DQC1}(z)\}_{z\in\{0,1\}}$.
Second, 
what Bob does in our protocol is only sending a single bit to Alice,
while the GES considers a more general setup: 
Bob sends Alice poly-length bit strings, and
multiple rounds of message exchanges are done between
Alice and Bob.
It is an open problem whether we can generalize our no-go result 
for more general setups (see Sec.~\ref{sec:discussion}). 
Third, our definition of the blindness given above is a minimum one,
and in fact it is a necessary condition for the more general
definition of the blindness in Refs.~\cite{AFK,ACGK}.
Our no-go result can be shown with any reasonable definition of the blindness 
as long as it includes the above definition of the blindness
as a necessary condition.
Finally, in our protocol, we have assumed that
a valid key is always obtained.
We can generalize it to the following:
on input $x$, the key generation algorithm $K$
outputs $(k,f)\leftarrow K(x)$. If $f=success$, $k$ is a valid key. 
If $f=fail$, $k$ is an invalid key.
The probability
that the key generation algorithm $K$ outputs $f=success$ is
at least $1/2+1/poly(|x|)$.
In this case, Alice has only to run $K$ until
she gets $f=success$.

\section{Result}
\label{sec:result}
The main result of the present paper is the following.

\begin{theorem}
\label{main}
If the above delegation protocol satisfies both the
$\epsilon$-correctness and blindness simultaneously with
$0\le\epsilon<1$,
then ${\rm NQP}\subseteq{\rm NP}/{\rm poly}$.
\end{theorem}

{\it Proof}.
Let $A=(A_{yes},A_{no})$ be a problem
in NQP. Then, from Theorem~\ref{thm:NQP_DQC1}, 
$A$ is in ${\rm NQP}_{\rm DQC1}$.
Therefore, there exists a polynomial-time uniformly generated
family $\{V_x\}_x$ of quantum circuits
such that
\begin{itemize}
\item
If $x\in A_{yes}$ then $p_{V_x}^{DQC1}(1)>0$.
\item
If $x\in A_{no}$ then $p_{V_x}^{DQC1}(1)=0$.
\end{itemize}
Let $s$ be a natural number. Let $k_s$ be any key for $1^s$,
i.e., $k_s\leftarrow K(1^s)$.
Let $a_s=E(1^s,k_s)$.
Bob sends Alice $b\in\{0,1\}$ with probability
$q_{a_s}(b)$ when he receives $a_s$ from Alice.
Let us consider the following
probabilistic algorithm
with advice.
\begin{itemize}
\item[1.]
On input $x$ with $|x|=s$,
receive $(a_s,q_{a_s}(0))$
as advice. Note that it is advice, because
both $a_s$ and $q_{a_s}(0)$ depend only on $s$.
(Note that we consider only quantum circuits whose
acceptance probabilities can be represented exactly in
poly-length bit strings, such as circuits consisting
of $H$ and Toffoli.)
\item[2.]
Run the key generation algorithm $K$ on
input $x$. 
Let $k$ be the obtained key, i.e., $k\leftarrow K(x)$.
Run the encryption algorithm $E$ on input $(x,k)$.
If
$E(x,k)\neq a_s$,
reject.
If $E(x,k)=a_s$, then 
generate $b\in\{0,1\}$ with probability
$q_{a_s}(b)$, and run the decryption algorithm $D$
on input $(x,k,b)$. Let $\xi=D(x,k,b)$,
where $\xi\in\{0,1\}$.
If $\xi=1$, accept.
If $\xi=0$, reject.
\end{itemize}
Because of the correctness, 
\begin{eqnarray*}
\big|{\rm Pr}(\xi=1)-p_{V_x}^{DQC1}(1)\big|&\le&\epsilon p_{V_x}^{DQC1}(1)
\end{eqnarray*}
for a certain $0\le\epsilon<1$.
The acceptance probability $p_{acc}$ 
of the above probabilistic algorithm with advice is 
\begin{eqnarray*}
p_{acc}
=\eta\times {\rm Pr}(\xi=1),
\end{eqnarray*}
where 
$\eta$ is the probability that the key generation algorithm $K$
on input $x$
outputs a key $k$ such that
$a_s=E(x,k)$.
Because of the blindness, $\eta>0$.
Hence, if $x\in A_{yes}$ then 
\begin{eqnarray*}
p_{acc}\ge \eta (1-\epsilon)p_{V_x}^{DQC1}(1)>0. 
\end{eqnarray*}
If $x\in A_{no}$ then 
\begin{eqnarray*}
p_{acc}\le \eta (1+\epsilon)p_{V_x}^{DQC1}(1)=0. 
\end{eqnarray*}
Therefore, $A$ is in ${\rm NP}/{\rm poly}$,
and we have shown ${\rm NQP}\subseteq{\rm NP}/{\rm poly}$.
\fbox

The consequence, ${\rm NQP}\subseteq{\rm NP}/{\rm poly}$,
leads to the collapse of the polynomial-time hierarchy
to the third level due to the following lemma.
Because the polynomial-time hierarchy is not believed to
collapse, Theorem~\ref{main} suggests
the impossibility of
the classical blind DQC1 sampling.

\begin{lemma}
If ${\rm NQP}\subseteq{\rm NP}/{\rm poly}$,
then the polynomial-time hierarchy collapses to the third level. 
\end{lemma}

{\it Proof}.
Note that
\begin{eqnarray*}
{\rm coNP}\subseteq{\rm PH}
\subseteq\widehat{\rm BP}\cdot{\rm coC}_={\rm P}
=\widehat{\rm BP}\cdot {\rm NQP}
\subseteq\widehat{\rm BP}\cdot {\rm NP}/{\rm poly}
\subseteq{\rm NP}/{\rm poly}.
\end{eqnarray*}
Here, the second inclusion is from Corollary 2.5 of Ref.~\cite{TO92}.
The third equality is from ${\rm coC}_={\rm P}={\rm NQP}$
of Ref.~\cite{FGHP99}.
The proof of the last containment, 
$\widehat{\rm BP}\cdot{\rm NP}/{\rm poly}\subseteq{\rm NP}/{\rm poly}$,
is similar to that of ${\rm BPP}\subseteq{\rm P}/{\rm poly}$
(see Appendix~\ref{app:BP}).
Finally, ${\rm coNP}\subseteq{\rm NP}/{\rm poly}$
leads to the collapse of the polynomial-time hierarchy to the
third level~\cite{Yap}.
(Actually, ${\rm coNP}\subseteq{\rm NP}/{\rm poly}$ leads
to the stronger result ${\rm PH}={\rm S}_2^{\rm NP}$~\cite{Cai},
where $\Sigma_2^p\cup\Pi_2^p\subseteq {\rm S}_2^{\rm NP}\subseteq
\Sigma_3^p$.)
\fbox

\section{Generalizations}
\label{sec:generalizations}
In this paper, we have shown that if a classical client can blindly delegate
the DQC1 sampling then the polynomial-time hierarchy collapses
to the third level.
It is clear that
the same result holds for 
another subuniversal model $M$ if
${\rm NQP}={\rm NQP}_M$ is satisfied, where
${\rm NQP}_M$ is the NQP whose quantum circuits
are restricted to the model $M$.
For example,
we can show the following:

\begin{theorem}
If the sampling of 
$\{p_V^{IQP,m}(z)\}_{z\in\{0,1\}}$
can be classically delegated satisfying both the
$\epsilon$-correctness and blindness simultaneously
with $0\le\epsilon<1$, then
${\rm coNP}\subseteq{\rm NP}/{\rm poly}$.
\end{theorem}

Furthermore, it is clear that the same result holds for
other subuniversal models $M$, such as
the Boson Sampling model~\cite{AA}, the depth-4 model~\cite{TD}, and
the random circuit model~\cite{random}, 
because ${\rm NQP}={\rm NQP}_M$ for these models.
(It is known that these models are universal under a postselection.
If we accept when the postselection is successful
and the original circuit accepts, then the acceptance probability
is proportional to the acceptance probability of a
universal circuit. See Appendix~\ref{app:NQP_IQP}.)

\section{Discussion}
\label{sec:discussion}
In this paper, we have considered the delegation protocol where Bob sends
only a single bit to Alice.
It is an open problem whether we can generalize
our no-go result for
more general delegation protocols.
For example, what happens if Bob sends a poly-length bit string to Alice?
(It is easy to see that Theorem~\ref{main} can be generalized to the 
delegation protocol
where Bob sends Alice a log-length bit string.)
Furthermore, what happens if multiple rounds of message exchanges are done
between Alice and Bob?

The argument used in the proof
of Theorem~\ref{main} does not seem to be 
directly applied to these generalized cases.
For example, let us modify our delegation protocol
in such a way that,
instead of the single bit,
Bob sends Alice a poly-length bit string $b\in\{0,1\}^{poly(|x|)}$
with probability $q_a(b)$ when he receives 
$a\in\{0,1\}^{poly(|x|)}$ from Alice.
Then we can show the following.

\begin{theorem}
If such a modified delegation protocol satisfies both the 
$\epsilon$-correctness
and blindness simultaneously with $0\le\epsilon<1$, 
then ${\rm NQP}\subseteq{\rm NP}/{\rm rpoly}$.
\end{theorem}

{\it Proof}.
Let $A=(A_{yes},A_{no})$ be a problem in NQP.
Then, from Theorem~\ref{thm:NQP_DQC1}, $A$ is in ${\rm NQP}_{\rm DQC1}$.
Therefore, there exists a polynomial-time uniformly generated
family $\{V_x\}_x$ of quantum circuits
such that
\begin{itemize}
\item
If $x\in A_{yes}$ then $p_{V_x}^{DQC1}(1)>0$.
\item
If $x\in A_{no}$ then $p_{V_x}^{DQC1}(1)=0$.
\end{itemize}
Let $s$ be a natural number. Let $k_s$ be any key for $1^s$,
i.e., $k_s\leftarrow K(1^s)$.
Let $a_s=E(1^s,k_s)$.
Bob sends Alice $b\in\{0,1\}^{poly(s)}$ with
probability $q_{a_s}(b)$ when he receives
$a_s$ from Alice.
Let us consider the following
probabilistic algorithm
with probabilistic advice.
\begin{itemize}
\item[1.]
On input $x$ with $|x|=s$,
receive $a_s$ and the probability distribution
$\{q_{a_s}(b)\}_{b\in\{0,1\}^{poly(s)}}$
as advice. Note that they are advices, because
both $a_s$ and $\{q_{a_s}(b)\}_{b\in\{0,1\}^{poly(s)}}$ depend only on $s$.
\item[2.]
Run the key generation algorithm $K$ on
input $x$. 
Let $k$ be the obtained key, i.e., $k\leftarrow K(x)$.
Run the encryption algorithm $E$ on input $(x,k)$.
If
$E(x,k)\neq a_s$,
reject.
If $E(x,k)=a_s$, then 
sample $b\in\{0,1\}^{poly(s)}$ from
$\{q_{a_s}(b)\}_{b\in\{0,1\}^{poly(s)}}$, 
and run the decryption algorithm $D$
on input $(x,k,b)$. Let $\xi=D(x,k,b)$,
where $\xi\in\{0,1\}$.
If $\xi=1$, accept.
If $\xi=0$, reject.
\end{itemize}
Because of the correctness, 
\begin{eqnarray*}
\big|{\rm Pr}(\xi=1)-p_{V_x}^{DQC1}(1)\big|&\le&\epsilon p_{V_x}^{DQC1}(1)
\end{eqnarray*}
for a certain $0\le\epsilon<1$.
The acceptance probability $p_{acc}$ 
of the above algorithm is 
\begin{eqnarray*}
p_{acc}
=\eta\times {\rm Pr}(\xi=1),
\end{eqnarray*}
where 
$\eta$ is the probability that the key generation algorithm $K$
on input $x$
outputs a key $k$ such that
$a_s= E(x,k)$.
Because of the blindness, $\eta>0$.
Hence, if $x\in A_{yes}$ then 
\begin{eqnarray*}
p_{acc}\ge\eta(1-\epsilon)p_{V_x}^{DQC1}(1)>0.
\end{eqnarray*}
If $x\in A_{no}$ then 
\begin{eqnarray*}
p_{acc}\le\eta(1+\epsilon)p_{V_x}^{DQC1}(1)=0.
\end{eqnarray*}
Therefore, $A$ is in ${\rm NP}/{\rm rpoly}$,
and we have shown ${\rm NQP}\subseteq{\rm NP}/{\rm rpoly}$.
\fbox

However, it can be shown that
${\rm NP}/{\rm rpoly}={\rm ALL}$
(for a proof, see Appendix~\ref{app:ALL})~\cite{Andrucomment}.
Therefore we cannot conclude any unlikely consequence,
such as the collapse of the polynomial-time hierarchy.

\acknowledgements
TM thanks A. Gheorghiu for sending his Ph.D. thesis
and answering to some questions.
TM is supported by JST PRESTO No.JPMJPR176A and 
JSPS Grant-in-Aid for Young Scientists (B)
No.JP17K12637.
HM is supported by
JSPS KAKENHI grants No. 26247016, No. 16H01705 and No. 16K00015.

\appendix
\section{Proof of Theorem~\ref{thm:NQP_DQC1}}
\label{app:NQP_DQC1}
\begin{figure}[htbp]
\begin{center}
\includegraphics[width=0.3\textwidth]{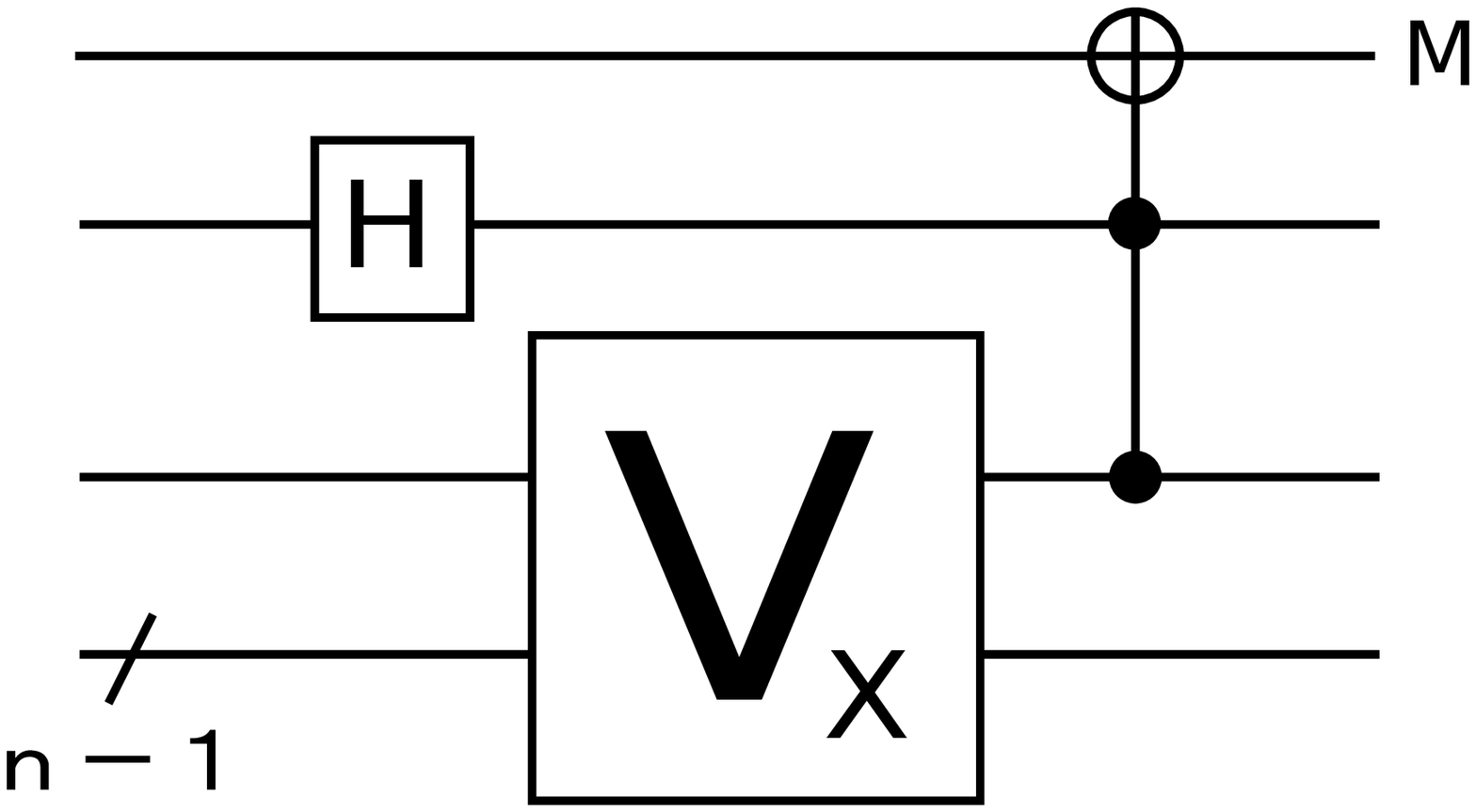}
\end{center}
\caption{
The circuit $W_x$.
$M$ means the computational-basis measurement.
The line with the slash $/$ means the set of $n-1$ qubits.
}
\label{Vprime}
\end{figure}

\begin{figure}[htbp]
\begin{center}
\includegraphics[width=0.65\textwidth]{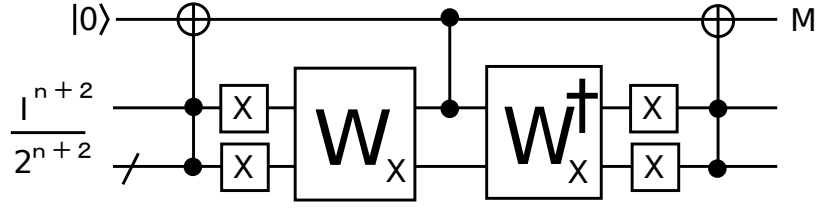}
\end{center}
\caption{The DQC1 model constructed from $W_x$.
The line with the slash $/$ means the set of $n+1$ qubits.
A gate acting on this line is applied on each qubit.
$M$ is the computational-basis measurement.
}
\label{DQC1}
\end{figure}

The inclusion ${\rm NQP}\supseteq{\rm NQP}_{\rm DQC1}$ is
trivial.
Let us show the other inclusion ${\rm NQP}\subseteq{\rm NQP}_{\rm DQC1}$.
Let $A=(A_{yes},A_{no})$ be a problem in NQP.
Then, there exists a polynomial-time uniformly generated family
$\{V_x\}_x$ of quantum circuits such that
\begin{itemize}
\item
If $x\in A_{yes}$ then $p_{V_x}(1)>0$.
\item
If $x\in A_{no}$ then $p_{V_x}(1)=0$.
\end{itemize}
Here, $p_{V_x}(1)\equiv\|(|1\rangle\langle1|\otimes I^{\otimes n-1})
V_x|0^n\rangle\|^2$.
Let us define the $(n+2)$-qubit circuit
$W_x$ as is shown in Fig.~\ref{Vprime}.
Then,
\begin{itemize}
\item
If $x\in A_{yes}$ then $0<p_{W_x}(1)<1$.
\item
If $x\in A_{no}$ then $p_{W_x}(1)=0$.
\end{itemize}
Here, $p_{W_x}(1)\equiv\|(|1\rangle\langle1|\otimes I^{\otimes n+1})
W_x|0^{n+2}\rangle\|^2$.
From $W_x$, we construct the DQC1 model
of Fig.~\ref{DQC1}.
By the straightforward calculation,
it is clear that
the probability $\tilde{p}$ of obtaining 1 when
the first qubit of the DQC1 model of Fig.~\ref{DQC1}
is measured in the computational basis
is
\begin{eqnarray*}
\tilde{p}=\frac{4p_{W_x}(1)(1-p_{W_x}(1))}{2^{n+2}}.
\end{eqnarray*}
Therefore, if $x\in A_{yes}$ then
$\tilde{p}>0$, and
if $x\in A_{no}$ then
$\tilde{p}=0$, which means that $A$ is in ${\rm NQP}_{\rm DQC1}$.
Hence we have shown ${\rm NQP}\subseteq{\rm NQP}_{\rm DQC1}$.

\section{Proof of Theorem~\ref{thm:NQP_IQP}}
\label{app:NQP_IQP}
Since ${\rm NQP}\supseteq{\rm NQP}_{\rm IQP}$ is trivial,
let us show
${\rm NQP}\subseteq{\rm NQP}_{\rm IQP}$.
Let $A=(A_{yes},A_{no})$ be a problem
in NQP. Then there exists a polynomial-time uniformly
generated family $\{V_x\}_x$ of 
quantum circuits such that
\begin{itemize}
\item
If $x\in A_{yes}$ then $p_{V_x}(1)>0$.
\item
If $x\in A_{no}$ then $p_{V_x}(1)=0$.
\end{itemize}
Here, $p_{V_x}(1)\equiv\|(|1\rangle\langle1|\otimes I^{\otimes n-1})
V_x|0^n\rangle\|^2$.
Since the IQP model is universal under a postselection,
for any $V_x$
there exist an IQP circuit $W_x$ and $s=poly(|x|)$ such that
\begin{eqnarray*}
\frac{(|1^s\rangle\langle1^s|\otimes I^{\otimes n})W_x|0^{n+s}\rangle}
{\sqrt{2^{-s}}}
=|1^s\rangle\otimes(V_x|0^n\rangle).
\end{eqnarray*}
Therefore 
\begin{eqnarray*}
p_{W_x}^{IQP,s+1}(1)=\Big\|(|1^{s+1}\rangle\langle1^{s+1}|\otimes 
I^{\otimes n-1})
W_x|0^{n+s}\rangle\Big\|^2
=\frac{p_{V_x}(1)}{2^s}.
\end{eqnarray*}
Therefore if $x\in A_{yes}$ then $p_{W_x}^{IQP,s+1}(1)>0$, and if
$x\in A_{no}$ then $p_{W_x}^{IQP,s+1}(1)=0$.
Hence $A$ is in ${\rm NQP}_{\rm IQP}$,
and we have shown ${\rm NQP}\subseteq{\rm NQP}_{\rm IQP}$.

\section{Proof of $\widehat{\rm BP}\cdot{\rm NP}/{\rm poly}
\subseteq{\rm NP}/{\rm poly}$}
\label{app:BP}
One way of showing it is to combine
Lemma 2.12 of Ref.~\cite{TO92},
$\widehat{\rm BP}\cdot{\rm K}\subseteq{\rm K}/{\rm poly}$,
with $({\rm K}/{\rm poly})/{\rm poly}\subseteq {\rm K}/{\rm poly}$.

Here, for the convenience of readers, we provide a direct proof.
Let $A=(A_{yes},A_{no})$ be a problem in
$\widehat{\rm BP}\cdot{\rm NP}/{\rm poly}$.
Then, for any polynomially bounded function 
$q:{\mathbb N}\to {\mathbb N}$, there exist a problem
$B=(B_{yes},B_{no})$ in ${\rm NP}/{\rm poly}$ and
a polynomially bounded function $r:{\mathbb N}\to{\mathbb N}$ such that
for every $x\in\{0,1\}^*$ it holds that
\begin{itemize}
\item
If $x\in A_{yes}$, then 
\begin{eqnarray*}
\big|\{z\in\{0,1\}^{r(|x|)}~|~\langle x,z\rangle\in B_{yes}\}\big|\ge
2^{r(|x|)}(1-2^{-q(|x|)}).
\end{eqnarray*}
\item
If $x\in A_{no}$, then 
\begin{eqnarray*}
\big|\{z\in\{0,1\}^{r(|x|)}~|~\langle x,z\rangle\in B_{no}\}\big|\ge
2^{r(|x|)}(1-2^{-q(|x|)}).
\end{eqnarray*}
\end{itemize}
Let us take $q(n)=n+1$.
For each $x\in A_{yes}\cap\{0,1\}^n$, 
the number of $z\in\{0,1\}^r$ such that $\langle x,z\rangle\notin B_{yes}$
is at most $2^{r(n)-q(n)}$.
For each $x\in A_{no}\cap\{0,1\}^n$, 
the number of $z\in\{0,1\}^r$ such that $\langle x,z\rangle\notin B_{no}$
is at most $2^{r(n)-q(n)}$.
Therefore, there exists at least one $z\in\{0,1\}^r$ such that
for all $x\in \{0,1\}^n$
\begin{itemize}
\item
If $x\in A_{yes}$ then $\langle x,z\rangle\in B_{yes}$.
\item
If $x\in A_{no}$ then $\langle x,z\rangle\in B_{no}$.
\end{itemize}
Let $\tilde{z}_n$ be such $z$.
Then, 
there exists an advice $\{\tilde{z}_n\}_n$ such that
for all $x\in \{0,1\}^*$
\begin{itemize}
\item
If $x\in A_{yes}$ then 
$\langle x,\tilde{z}_{|x|}\rangle\in B_{yes}$.
\item
If $x\in A_{no}$ then 
$\langle x,\tilde{z}_{|x|}\rangle\in B_{no}$.
\end{itemize}
Since $B=(B_{yes},B_{no})$ is in ${\rm NP}/{\rm poly}$, 
there exists a problem $C=(C_{yes},C_{no})$ in NP and
advice $\{\eta_n\}_n$
such that
\begin{itemize}
\item
If $y\in B_{yes}$ then $\langle y,\eta_{|y|}\rangle\in C_{yes}$. 
\item
If $y\in B_{no}$ then $\langle y,\eta_{|y|}\rangle\in C_{no}$. 
\end{itemize}
Therefore,
\begin{itemize}
\item
If $x\in A_{yes}$ then 
$\langle x,\tilde{z}_{|x|},\eta_{|\langle x,\tilde{z}_{|x|}\rangle|}
\rangle \in C_{yes}$.
\item
If $x\in A_{no}$ then 
$\langle x,\tilde{z}_{|x|},\eta_{|\langle x,\tilde{z}_{|x|}\rangle|}
\rangle \in C_{no}$.
\end{itemize}
Hence we have shown
$A$ is in ${\rm NP}/{\rm poly}$.

\section{Proof of ${\rm NP}/{\rm rpoly}={\rm ALL}$}
\label{app:ALL}
Here we show
${\rm NP}/{\rm rpoly}={\rm ALL}$. The proof is 
essentially the same as that of 
${\rm PP}/{\rm rpoly}={\rm ALL}$~\cite{Aar05}. 

Let $A=(A_{yes},A_{no})$ be any problem.
Let $f:\{0,1\}^*\to\{0,1,\perp\}$ be a function such
that $f(x)=1$ if and only if $x\in A_{yes}$,
and
$f(x)=0$ if and only if $x\in A_{no}$.
Let $q_s:\{0,1\}^s\times \{0,1,\perp\}\to[0,1]$ be the
probability distribution such that
\begin{eqnarray*}
q_s(x,y)=
\left\{
\begin{array}{ll}
\frac{1}{2^s}&y=f(x)\\
0&y\neq f(x)
\end{array}
\right.
\end{eqnarray*}
for all $(x,y)\in\{0,1\}^s\times\{0,1,\perp\}$. 
Let us consider the following probabilistic algorithm with 
probabilistic advice:
\begin{itemize}
\item[1.]
On input $x$ with $|x|=s$, 
receive the probability distribution $q_s$ as advice.
\item[2.]
Sample $(x',y)$ from $q_s$.
If $x'\neq x$, reject.
If $x'=x$, see $y$.
If $y=1$, accept.
If $y\neq 1$, reject.
\end{itemize}
If $x\in A_{yes}$, the acceptance 
probability $p_{acc}$ is 
\begin{eqnarray*}
p_{acc}=\frac{1}{2^s}\times 1>0.
\end{eqnarray*}
If $x\in A_{no}$, the acceptance 
probability $p_{acc}$ is 
\begin{eqnarray*}
p_{acc}=\frac{1}{2^s}\times 0=0.
\end{eqnarray*}
Therefore, $A$ is in ${\rm NP}/{\rm rpoly}$.


\begin{thebibliography}{00}
\bibitem{error}
Theorem 3 of Ref.~\cite{ACGK}, 
which roughly says that
``if a classical blind delegation of
Boson Sampling is possible, then the polynomial-time hierarchy
collapses to the fourth level",
contains an error~\cite{private}.
On the other hand, another result, which roughly says 
that
``if a classical blind delegation of 
Boson Sampling is possible,
then the permanent can be computed with a small size 
circuit" is shown in Ref.~\cite{AndruPhD}.
\bibitem{Andrucomment}
One might think that this seems strange as 
${\rm MA}/{\rm rpoly}={\rm NP}/{\rm poly}$~\cite{Aar06} while NP is in MA.
However, adding randomized (or quantum) advice may change the inclusion 
order of the original complexity classes.
In our case, the definition of NP based on a probabilistic Turing machine 
(the unbounded-error acceptance criterion) 
is crucial to make ${\rm NP}/{\rm rpoly}={\rm ALL}$.
\end{thebibliography}
\end{document}